\input{aipcheck}

\documentclass[final]{aipproc}

\layoutstyle{6x9}

\def\mic{$\mu {\rm m}$}
\def\NLF{$233$ }

\usepackage{aas_macros}

\begin{document}

\title{Can the Near-IR Fluctuations Arise from Known Galaxy Populations?}


\author{Kari Helgason}{
  address={Department of Astronomy, University of
Maryland, College Park, MD 20742, USA}
}

\author{Massimo Ricotti}{
  address={Department of Astronomy, University of
Maryland, College Park, MD 20742, USA}
}

\author{Alexander Kashlinsky}{
  address={ Observational Cosmology Laboratory, Code 665, NASA Goddard Space
Flight Center, Greenbelt MD 20771}
}

\begin{abstract}

Spatial Fluctuations in the Cosmic Infrared Background have now been measured out to sub-degree scales showing a strong clustering signal from unresolved sources. We attempt to explain these measurement by considering faint galaxy populations at z$<$6 as the underlying sources for this signal using \NLF measured UV, optical and NIR luminosity functions (LF) from a variety of surveys covering a wide range of redshifts. We populate the lightcone and calculate the total emission redshifted into the near-IR bands in the observer frame and recover the observed optical and near-IR galaxy counts to a good accuracy. Using a halo model for the clustering of galaxies with an underlying $\Lambda$CDM density field, we find that fluctuations from known galaxy populations are unable to account for the large scale CIB clustering signal seen by {\it HST}/NICMOS, {\it Spitzer}/IRAC and {\it AKARI}/IRC and continue to diverge out to larger angular scales. Our purely empirical reconstruction shows that known galaxy populations are not responsible for the bulk of the fluctuation signal seen in the measurements and suggests an unknown population of very faint and highly clustered sources dominating the signal.

\end{abstract}

\maketitle

\begin{figure}
  \includegraphics[width=.7\textheight]{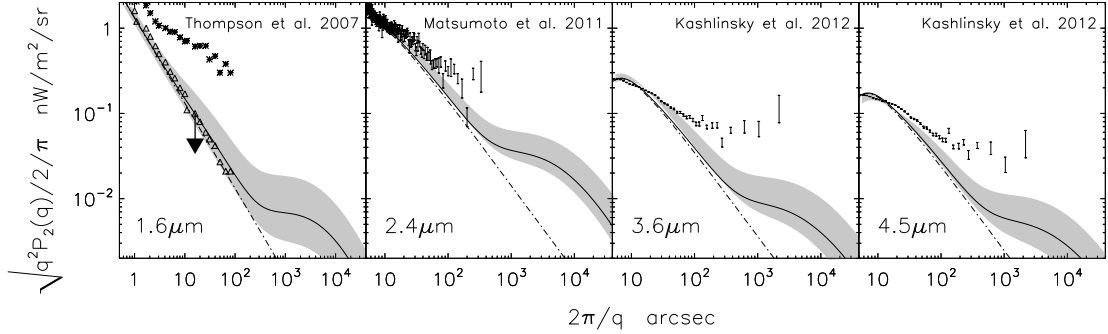}
  \caption{Models of the unresolved near-IR fluctuations compared to measurements from authors listed in the panels. Here, we approximate the root-mean-square fluctuations as $(q^2P_2(q)/2\pi)^{1/2}\!\sim\!\langle \delta F_{\mathbf{\theta}}^2\rangle^{1/2}$ \citep{Kashreview}. We have normalized the models to the shot noise levels (dot-dashed lines) reached in these studies. The solid curves show the total contribution from clustering and shot noise whereas the light shaded areas indicate the region bracketed by our two faint-end scenarios. }
  \label{fig_fluctuations}
\end{figure}

A significant clustering signal has been tentatively detected in the near-IR background fluctuations after removing resolved sources down to faint levels. Whereas current measurements from {\it HST}/NICMOS \citep{Thompson07a}, {\it AKARI}/IRC \citep{Matsumoto11} and {\it Spitzer}/IRAC \citep{Kashlinsky12} are mutually consistent, the nature of the sources producing the observed signal is debated. \citet{Kashlinsky12} attribute their large scale signal to high-z sources whereas \citet{Thompson07a} argue that their colors are consistent with normal galaxy populations at z$<$8. Identifying the origins of the unresolved fluctuations in the CIB requires good understanding of the resolved foreground galaxies their unresolved counterparts.
\begin{figure}
  \includegraphics[height=.32\textheight]{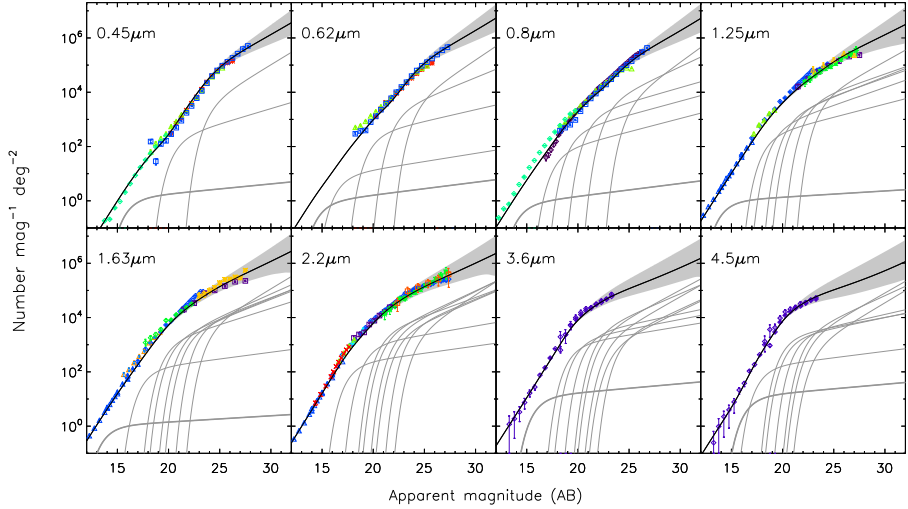}
  \caption{Galaxy number counts associated with our empirical model our default description (solid curve) including the regions of bracketed by our two faint-end scenarios (gray shaded areas). The gray curves show the underlying template LFs which we interpolate and integrate to obtain the number counts. The low-$z$ LF dominate the bright counts whereas high- and intermediate redshift LFs dominate the faint counts (from left to right). The most local available LF is shown as thick gray curves to demonstrate their negligible contribution to the faint counts.}
  \label{fig_nc}
\end{figure}
In \citet{Helgason12}, we consider the possibility of CIB fluctuations arising from low and intermediate redshift galaxies beyond the detection threshold of current NIR observatories. The total emission seen in the near-IR bands ({\small $JHKLM$}) depends on the contribution of local near-IR galaxies as well as redshifted light radiated at shorter rest-frame wavelengths. To quantify the present day background produced by galaxies, we have utilized measurements of luminosity functions probing all {\em rest-frame} wavelengths in the interval 0.1$<\! \lambda \!<$5.0\mic\ anywhere in the redshift cone. This results in a compilation of \NLF LFs from a large variety of surveys shown in Table 1 of \citet{Helgason12}. Our approach therefore predicts the levels of CIB fluctuations directly from the available data, assuming only i) standard $\Lambda$CDM model of structure formation and ii), the validity of a Schechter-type LF after fitting its parameters to the data. All the LFs we use have been characterized by a Schechter function and we explore the evolution of the population in terms of measured Schechter parameters at different redshifts,
\begin{equation} \label{eqn:poplight}
  \Phi(M|z) = 0.4 \ln{(10)}\phi^{\star}(z) \left(10^{0.4(M^{\star}(z)-M)}\right)^{\alpha(z) + 1} \exp{(-10^{0.4(M^{\star}(z)-M)})}
\end{equation}
where $\phi^{\star}$ is the normalization, $M^{\star}$ is the characteristic absolute magnitude and $\alpha$ is the faint-end slope. Providing fitting formulae for the multi-band evolution of these parameters, we populate the lightcone in the $z=0-6$ range and calculate the total emission redshifted into the near-IR bands in the observer frame, recovering the galaxy number counts in the 0.45-4.5\mic\ range (see Figure~\ref{fig_nc}). Because of the uncertainty in the faint-end slope of the LF, we consider two scenarios for the amount of steepening of $\alpha$ with redshift, 1) moderate evolution with a LF cutoff at 10$^{-4}L^\star$, and 2) strong evolution with a LF cutoff at 10$^{-8}L^\star$, within the range allowed by deep counts data (see shaded regions in Figure~\ref{fig_nc}). 
We adopt a halo model \citep{Cooray&Sheth02} to describe the underlying clustering of galaxies which on large scales reproduces the linear $\Lambda$CDM density field multiplied by a bias factor, $P_3(k) = b^2P^{lin}(k)$. The angular power spectrum of galaxies projected onto the sky can be related to their flux and evolving 3D power spectrum, $P_3(k)$, by the Limber approximation (for $\theta\ll$1 rad),
\begin{equation} \label{limber}
  P(q) = \frac{1}{c} \int \left[\frac{d\mathcal{F}}{dz}\right]^2 \frac{P_3(qd_A^{-1};z)}{ \frac{dt}{dz}d_A^2(z)} \frac{dz}{1+z},
\end{equation}
where $d_A$ is the comoving angular diameter distance and the quantity in the square brackets is the flux production rate derived empirically from our lightcones. The total power consists of both the clustering of galaxies (Eqn.~\ref{limber}) and shot noise from unresolved sources, $P_{SN} = \int dm f(m)^2 dn/dm$. Figure~\ref{fig_fluctuations} shows our computed fluctuations in the unresolved CIB and compares the results to available measurements illustrating that these sources are unable to account for the bulk of the CIB clustering signal seen by {\it HST}/NICMOS, {\it Spitzer}/IRAC and {\it AKARI}/IRC out to angular scales $\sim$20 arcmin. We interpret this as evidence for a contribution from a hitherto unknown population.

Both \citet{Matsumoto11} and \citet{Kashlinsky12} present evidence against the signal originating in Galactic or Solar system foreground emissions. Since we found that known galaxy populations (extrapolated to very faint limits) cannot explain the measurements, the clustering signal seen in the CIB fluctuations must originate in new populations so far unobserved in galaxy surveys. \citet{KAMM4} demonstrate the lack of correlations between the ACS maps with sources down to AB mag of $\simeq$28, and the source-subtracted {\it Spitzer}/IRAC maps. This implies that either the CIB fluctuations originate in a large unknown population of very small systems at low/intermediate redshifts, or they are produced by high redshift, z$>$7, populations whose Lyman break (at rest 0.12\mic) is redshifted past the longest ACS wavelength (at 0.9\mic).

\bibliographystyle{aipproc}   


\end{document}